\documentclass[twocolumn,showpacs,superscriptaddress,preprintnumbers,amsmath,amssymb,aps]{revtex4}

\usepackage{graphicx}
\usepackage{dcolumn}
\usepackage{bm}

\begin{document}

\newcommand\hatbfn{{\bf\hat{n}}}

\preprint{astroph/0307335}

\title{Physical Evidence for Dark Energy}

\author{Ryan Scranton}
\email{scranton@bruno.phyast.pitt.edu}
\affiliation{University of Pittsburgh, Department of Physics and 
Astronomy, Pittsburgh, PA 15260, USA}
\author{Andrew J. Connolly}
\affiliation{University of Pittsburgh, Department of Physics and 
Astronomy, Pittsburgh, PA 15260, USA}
\author{Robert C. Nichol}
\affiliation{Department of Physics, Carnegie Mellon University, Pittsburgh, 
PA 15213, USA}
\author{Albert Stebbins}
\affiliation{NASA/Fermilab Astrophysics Center, Batavia, IL 60510, USA}
\author{Istv\'an Szapudi}
\affiliation{Institute for Astronomy, University of Hawaii, Honolulu, 
HI 96822, USA}
\author{Daniel J. Eisenstein}
\affiliation{Steward Observatory, University of Arizona, Tucson, AZ, 
85721, USA}
\author{Niayesh Afshordi}
\affiliation{Princeton University Observatory, Princeton, NJ 08544, USA}
\author{Tamas Budavari}
\affiliation{Department of Physics and Astronomy, The Johns Hopkins 
University, Baltimore, MD 21218, USA}
\author{Istv\'an Csabai}
\affiliation{Department of Physics, E\"{o}tv\"{o}s University, Budapest,
Pf.\ 32, Hungary, H-1518}
\author{Joshua A. Frieman}
\affiliation{Astronomy and Astrophysics Department, University of 
Chicago, Chicago, IL 60637, USA}
\affiliation{NASA/Fermilab Astrophysics Center, Batavia, IL 60510, USA}
\author{James E. Gunn}
\affiliation{Princeton University Observatory, Princeton, NJ 08544, USA}
\author{David Johnston}
\affiliation{Center for Cosmological Physics, University of Chicago, 
Chicago, IL 60637, USA}
\affiliation{Astronomy and Astrophysics Department, University of 
Chicago, Chicago, IL 60637, USA}
\author{Yeong-Shang Loh}
\affiliation{Princeton University Observatory, Princeton, NJ 08544, USA}
\author{Robert H. Lupton}
\affiliation{Princeton University Observatory, Princeton, NJ 08544, USA}
\author{Christopher J. Miller}
\affiliation{Department of Physics, Carnegie Mellon 
University, Pittsburgh, PA 15213, USA}
\author{Erin Scott Sheldon}
\affiliation{Center for Cosmological Physics, University of Chicago, 
Chicago, IL 60637, USA}
\author{Ravi K. Sheth}
\affiliation{University of Pittsburgh, Department of Physics and 
Astronomy, Pittsburgh, PA 15260, USA}
\author{Alexander S. Szalay}
\affiliation{Department of Physics and Astronomy, The Johns Hopkins 
University, Baltimore, MD 21218, USA}
\author{Max Tegmark}
\affiliation{Department of Physics and Astronomy, University of Pennsylvania,
Philadelphia, PA 19104, USA}
\author{Yongzhong Xu}
\affiliation{Department of Physics and Astronomy, University of Pennsylvania,
Philadelphia, PA 19104, USA}
\author{Scott F. Anderson}
\affiliation{Department of Astronomy, University of Washington, Seattle, WA 
98195, USA}
\author{Jim Annis}
\affiliation{NASA/Fermilab Astrophysics Center, Batavia, IL 60510, USA}
\author{John Brinkmann}
\affiliation{Apache Point Obs., Sunspot, NM 88349-0059, USA}
\author{Neta A. Bahcall}
\affiliation{Princeton University Observatory, Princeton, NJ 08544, USA}
\author{Robert J. Brunner}
\affiliation{Department of Astronomy, University of Illinois, 
Urbana, IL 61801, USA}
\author{Masataka Fukugita}
\affiliation{Institute for Cosmic Ray Research, University of Tokyo, Kashiwa 
City, Japan}
\author{\v{Z}eljko Ivezi\'{c}}
\affiliation{Princeton University Observatory, Princeton, NJ 08544, USA}
\affiliation{H.N. Russell Fellow}
\author{Stephen Kent}
\affiliation{NASA/Fermilab Astrophysics Center, Batavia, IL 60510, USA}
\author{Don Q. Lamb}
\affiliation{Astronomy and Astrophysics Department, University of 
Chicago, Chicago, IL 60637, USA}
\author{Brian C. Lee}
\affiliation{Lawrence Livermore National Laboratory, Livermore, CA 94550, USA}
\author{Jon Loveday}
\affiliation{Sussex Astronomy Centre, University of Sussex, Falmer, 
Brighton BN1 9QJ, UK}
\author{Bruce Margon}
\affiliation{Space Telescope Institute, Baltimore, MD 21218, USA}
\author{Timothy McKay}
\affiliation{University of Michigan, Department of Physics, 
Ann Arbor, MI 48109}
\author{Jeffrey A. Munn}
\affiliation{US Naval Observatory, Flagstaff Station, Flagstaff, AZ 
86002-1149, USA}
\author{David Schlegel}
\affiliation{Princeton University Observatory, Princeton, NJ 08544, USA}
\author{Donald P. Schneider}
\affiliation{Department of Astronomy and Astrophysics, The Pennsylvania 
State University, University Park, PA, 16802, USA}
\author{Chris Stoughton}
\affiliation{NASA/Fermilab Astrophysics Center, Batavia, IL 60510, USA}
\author{Michael S. Vogeley}
\affiliation{Department of Physics, Drexel University, Philadelphia, PA 19104}

\date{\today}

\begin{abstract}

We present measurements of the angular cross-correlation between
luminous red galaxies from the Sloan Digital Sky Survey and the cosmic
microwave background temperature maps from the Wilkinson Microwave
Anisotropy Probe.  We find a statistically significant achromatic
positive correlation between these two data sets, which is consistent
with the expected signal from the late Integrated Sachs-Wolfe (ISW)
effect.  We do not detect any anti-correlation on small angular scales
as would be produced from a large Sunyaev-Zel'dovich (SZ) effect,
although we do see evidence for some SZ effect for our highest
redshift samples.  Assuming a flat universe, our preliminary detection
of the ISW effect provides independent physical evidence for the
existence of dark energy.

\end{abstract}

\pacs{98.65.Dx,98.62.Py,98.70.Vc,98.80.Es}
\maketitle

\section{\label{sec:intro}Introduction}

As photons from the Cosmic Microwave Background (CMB) travel to us
from the surface of last scattering, 
they can experience a number of physical processes. These include the
Sunyaev-Zel'dovich (SZ\citep{ZeldovichSunyaev1967}) effect, which is
the inverse Compton scattering of CMB photons by a hot ($T > 10^7$ K)
ionized gas\citep{Birkinshaw1999}, and the late Integrated
Sachs--Wolfe (ISW\citep{SachsWolfe1967,HuDodelson2002}) effect, which
is the integrated differential gravitational redshift caused by the
evolution of gravitational potentials along the path traveled by the
photons.  At frequencies less than $217$ GHz, the SZ effect produces
an anti-correlation between the temperature of the CMB and galaxies on
small angular scales. The ISW effect generates a weak positive
correlation at large angular scales for all
frequencies\citep{PeirisSpergel2000}. For a universe with a flat
geometry\citep{SpergelEtAl2003}, the detection of any ISW effect
provides strong physical evidence for the existence of dark energy,
and can be used to measure its equation of state
\citep{Coble1997,Caldwell98,Hu98,CorasanitiEtAl2003}. In this paper,
we attempt the detection of the ISW and SZ effects through the
cross-correlation of new, large area, clean galaxy and CMB maps of the
sky \citep{PeirisSpergel2000,Cooray2002a}, which reduce the errors due
to sample variance on large scales
and the Poisson error on small scales relative to that of previous
attempts\citep{boughn/crittenden:2003,fosalba/etal:2003a,Myers2003,Nolta2003}.

\section{Data \& Estimators} \label{sec:data} 

\begin{figure}[t]
\includegraphics[width=240pt]{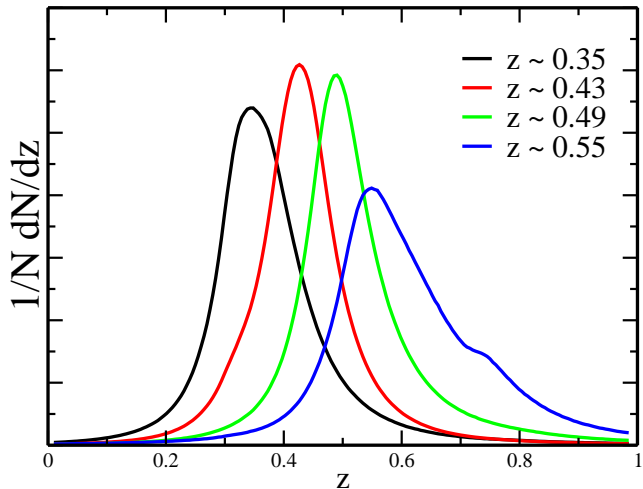}
\caption{\label{fig:lrg_dndz} The photometric redshift distributions
for the four LRG subsamples, taking into account the covariance
between photometric redshift and galaxy type.  To select these
galaxies, we imposed a color cut of $0.7(g - r) + 1.2((r - i) - 0.18)
> 1.6$ and $(g - r) > 1$.  We then define $d_\perp = (r - i) - (g -
r)/8$, and step through $d_\perp$ in increments of 0.2 starting at
$0.2 < d_\perp < 0.4$ and ending at $d_\perp > 0.8$ to determine our
four redshift samples, from lowest to highest redshift respectively.
There are 0.4 million LRGs in the $z \sim 0.35$ subsample, 0.8 million
in the $z \sim 0.43$ subsample, 1 million in the $z \sim 0.49$
subsample, and 0.7 million in the $z \sim 0.55$ subsample.}
\end{figure}

The parent SDSS galaxy data used herein is nearly 25 million galaxies
taken from $\simeq3400$ square degrees of the sky imaged by the Sloan
Digital Sky Survey
(SDSS\citep{Fukugita1996,Gunn1998,YorkEtAl2000,Abazajian2003,
Hogg2001,Pier2003,Smith2002,Stoughton2002}). We apply a flux limit of 
$i<21$ and regions with poor image quality have been excluded from our 
analysis ({\it i.e.}, data with a point--spread function of worse than 
1.5 arcseconds).  From this basic data set, we have constructed four 
different subsets of galaxies using both the Luminous Red Galaxy (LRG) 
selection method\citep{Eisenstein2001} and a photometric redshift 
estimate for every galaxy\citep{Budavari2000}.  As shown in 
Figure~\ref{fig:lrg_dndz}, these subsamples span a range
of redshift from $z\sim0.3$ to $z\sim0.8$, providing an efficient
method for tracing the large--scale distribution of galaxies in the
Universe as a function of redshift.

For the CMB data, we use the 3 primary combined-channel maps (Q, V, W) from
the Wilkinson Microwave Anisotropy Probe (WMAP\citep{SpergelEtAl2003}), and
the ``clean'' map of Tegmark et al.\citep{TegmarkEtAl2003}.  We also apply a
mask to exclude foreground contamination from the Galaxy (kp12 for the V \& W
bands and kp2 for the Q band).  To isolate the ISW effect, we also generate a
smoothed version of the ``clean'' map, by applying a Gaussian kernel with
a FWHM of one degree to remove CMB primary anisotropies on small angles.

The natural pixelization scheme for the WMAP maps is HEALPix 
\footnote{\rm http://www.eso.org/science/healpix} while for the
galaxies it is
SDSSPix\footnote{http://lahmu.phyast.pitt.edu/$\sim$scranton/SDSSPix}.
The latter is a new, hierarchical equal-area pixelization scheme
developed specifically for the SDSS geometry and fast correlation
statistics.  For our primary calculations we use the higher resolution
SDSSPix for accurate weighting, while for the extensive Monte Carlo
simulations discussed below, we use HEALPix to take advantage of the
SpICE\citep{szapudi} for fast cross--correlation
estimations. Both methods required re--pixelization of the original
maps. We have tested our re-pixelization methods by comparing the
auto-correlation functions of the original and re--sampled maps and we
find identical results and errors.

For the cross-correlation measurement, we use a simple pixel-based estimator.  
For each pixel $i$ in our map, we determine the galaxy and temperature 
over-densities ($\Delta^{\rm g}$ \& $\Delta^{\rm T}$, respectively),
\begin{equation}
\Delta^{\rm g}_i
          =\frac{n^{\rm g}_i-\overline{n^{\rm g}}}{\overline{n^{\rm g}}_i},
\qquad
\Delta^{\rm T}_i= T_i-\overline{T},
\label{eq:Delta_defs}
\end{equation}
where $T_i$ is the temperature in pixel $i$, $\overline{T}$ is the
mean temperature, $n^{\rm g}_i$ is the total number of galaxies in
pixel $i$, and $\overline{n^{\rm g}}_i$ is expected number of galaxies
in pixel $i$.  In addition, we can optimally weight each pair of
pixels to account for the variation of signal-to-noise (S/N) in the
pixels. Therefore, the two--point cross-correlation function at
angular separation $\theta_\alpha$ is given by
\begin{equation}
w_{\rm gT}(\theta_\alpha) = 
\frac{\sum_{i,j}\Delta^{\rm g}_i\,\Delta^{\rm T}_j\,f_i\,s_j\,
                                                          \Theta^\alpha_{ij}}
     {\sum_{n,m} f_{n} s_{m} \Theta^\alpha_{n,m}}\ ,
\label{eq:estimator}
\end{equation}
where $\theta_\alpha$ is the central value of an angular correlation
bin, $f_i$ is the fraction of pixel $i$ unmasked, $s_i$ is the inverse
of the noise in the WMAP pixel, and $\Theta^\alpha_{ij}$ is unity if
the separation between the pixels is within this angular bin and zero
otherwise.  On large angular scales the optimal CMB weighting is
uniform ({\it i.e.}, $s_i=1$, except where masked) as the signal is
dominated by sample variance rather than Poisson noise. For
completeness, we calculate the cross-correlation functions for both
the uniform and noise--weighted schemes.

To estimate the covariance $C(\theta_\alpha,\theta_\beta)$, we use two 
methods: jack--knife errors and random CMB maps.  For the former, we
divide our total area into $N_J$ equal-area subsamples and use the
jack--knife estimator\citep{scranton2002},
\begin{eqnarray}    
C_J(\theta_\alpha,\theta_\beta)&=&\frac{N_J-1}{N_J} \sum^{N_J}_{i=1} 
(\overline{w_{\rm gT}}(\theta_\alpha) - w_{\rm gT,i}(\theta_\alpha))\cr
&&\times(\overline{w_{\rm gT}}(\theta_\beta) - w_{\rm gT,i}(\theta_\beta)), 
\label{eq:jackknife}
\end{eqnarray}
where $\overline{w_{\rm gT}}(\theta)$ is the mean of $w_{\rm
gT}(\theta)$ for the $N_J$ measurements and $w_{{\rm gT},i}(\theta)$
is the measurement of the cross-correlation excluding the $i$th
subsample.  In order to produce a stable estimation of the covariance
matrix (non-singular and positive definite), we set $N_J = 2N_\theta$,
where $N_\theta$ is the number of angular bins.  Tests with different
values of $N_J$ near to $2N_\theta$ generally produced values of
$\chi^2$ (see Section \ref{sec:significance}) within 15\% of the values
obtained using $N_J = 2N_\theta$.

While the jack--knife method gives a good approximation of the sample
variance, it is less sensitive to variance on angular scales larger
than the survey size.  For a more conservative estimate of the
covariance, we generated 200 random CMB maps based on the angular
power spectrum of the ``clean'' map and 200 random CMB maps based on
the angular power spectrum of our smoothed WMAP map.  These maps
were cross-correlated against the four galaxy maps (using
SpICE\citep{szapudi}) and the results were re-sampled into the same
angular bins as used in the SDSSPix analysis.  The covariance matrix
is then
\begin{eqnarray}    
C_R(\theta_\alpha,\theta_\beta)&=&\frac{1}{N} \sum^{N=200}_{i=1} 
(\overline{w_{\rm gT}}(\theta_\alpha) - w_{\rm gT,i}(\theta_\alpha))\cr
&&\times(\overline{w_{\rm gT}}(\theta_\beta) - w_{\rm gT,i}(\theta_\beta)). 
\label{eq:random}
\end{eqnarray}

\section{Results} \label{sec:results}

\begin{figure}[t]
\includegraphics[width=240pt]{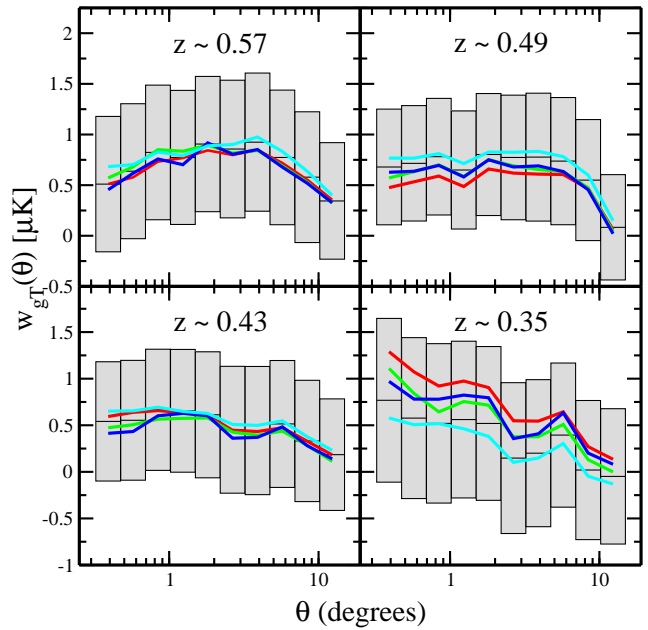}
\caption{The uniform-weighted galaxy--WMAP cross-correlation functions for the
four galaxy subsamples.  The shaded boxes show the measurement and the
$1\sigma$ jack--knife errors for the ``clean'' galaxy--WMAP cross-correlation
function. The errors on the other functions are nearly identical.  The Q map
is given by the red line, V by the green line, W by the blue line and 
smoothed map by the cyan line.}
\label{fig:lrg_w_gt}
\end{figure}

We present the cross--correlation functions for the four redshift
subsamples and the four WMAP maps in Figure~\ref{fig:lrg_w_gt}.  For
the three highest redshift galaxy subsamples, we find a consistent
non--zero signal. We also find that the cross--correlation signal is
achromatic in the Q, V, W frequency maps. We see little evidence for
an anti-correlation between the galaxy density and CMB temperature on
the smallest angular scales, as expected from the SZ effect.  For the
lowest redshift subsample, we see a different shape to the
cross--correlation function and all the angular bins are consistent
with zero within their errors.  This lowest redshift subsample also
shows a lack of agreement between the Q, V, and W frequency maps.

\section{Discussion} \label{sec:discussion}

\subsection{Systematic Error Tests} \label{sec:systematics}

Our primary concern is the possible contamination of the CMB signal by
residual Galactic synchrotron or dust emission and the galaxy catalogs by
Galactic stars.  This could result in a non--zero correlation between the maps.
To check against this, we calculate the stellar over-density in each pixel,
using the same color cuts as for the four galaxy subsamples discussed above,
and substitute these over-densities into Equation~\ref{eq:estimator}.  For the
three highest redshift subsamples, we find flat, featureless star-WMAP
cross-correlation functions consistent with zero at the $1\sigma$ level.  For
the lowest redshift sample ($z\sim 0.35$), the star-WMAP cross-correlation
function is nearly identical to the galaxy--WMAP measurement, suggesting
significant contamination.

The cross--correlation functions for the three highest redshift galaxy
subsamples are achromatic in the WMAP Q, V and W frequency maps. In
addition to the visual agreement in Figure~\ref{fig:lrg_w_gt}, we
cross-correlated the three highest redshift galaxy subsamples against
four linear combinations of the Q, V, and W maps. These combination
were designed to detect achromatic effects (${\rm Q-V}$ and ${\rm V-W}$) and
possible contamination from synchrotron (${\rm (9Q-12V+3W)/16}$) and dust
emission (${\rm (3Q-12V+9W)/16}$). In all 12 cross-correlation functions,
the signal was suppressed relative to our original galaxy--WMAP signal
by a factor of $\sim\!10$ on all angular scales with errors consistent
with Poisson noise from the galaxies. Likewise, the shapes of the
cross-correlation functions were dissimilar to those seen for the
galaxy--WMAP cross-correlation functions.  These observations confirm
that our galaxy--WMAP measurements are not due to contamination by
synchrotron or dust emission from the Galaxy, or synchrotron emission
the host galaxies themselves\citep{Myers2003}, {\it i.e.,} over the
frequency range probed by WMAP, the intensity of any synchrotron
emission should decrease by an order of magnitude from Q to W. 

\subsection{Significance Tests} \label{sec:significance}

As with any angular correlation measurement, the individual angular
bins are highly correlated.  Therefore, we need to consider the the
full covariance matrix ($C_{R,J}$) when checking the statistical
significance of our measurements against the null hypothesis of a zero
cross--correlation function. Our $\chi^2$ values are given by
\begin{equation}
\chi_{R,J}^2= w_{\rm gT} C_{R,J}^{-1} w_{\rm gT}, 
\label{eq:chi2}
\end{equation}
where $C_{R,J}$ is the covariance matrix derived from either
Eqn~\ref{eq:random} (random maps) or Eqn~\ref{eq:jackknife}
(jack--knife errors) respectively. The $\chi^2$ values for each of our
uniform weighted cross--correlation functions, including the
smoothed map, are given in Table~\ref{tab:chi2} using both
covariance matrices. 
These values are in reasonable agreement with other measurements of
similar cross-correlation functions at larger angles
\citep{boughn/crittenden:2003,Nolta2003}.  The choice of weighting for the CMB
pixels does not change our results significantly.

For the three highest redshift samples, we exclude the null hypothesis
at $\ge 90\%$ confidence for every WMAP band and the ``clean'' and
smoothed maps using the jack--knife errors.  For the random map
errors, we can exclude the null hypothesis at $\ge 90\%$ for only the
two highest redshift samples (excluding the Q band and smoothed
map).  Excluding angular bins on scales $< 2$ degrees increases
the significance of the smoothed detection with the random map
errors but not to the 90\% confidence threshold.  Checking the number
of random maps with $\chi^2$ greater than the values in
Table~\ref{tab:chi2} verifies that our confidence estimates are
correct.

\begin{table}[t]
\begin{center}
\caption{The $\chi^2$ values derived from both covariance matrices
(jack--knife errors given in parentheses) for $N_\theta = 10$ angular bins}
\label{tab:chi2}
\begin{tabular}{@{\extracolsep{\fill}}c||c|c|c}
Subsample & Q & V & W \\
\hline
$z \sim 0.55$ & 10.4 (16.3) & 18.6 (20.4) & 35.7 (25.3) \\
$z \sim 0.49$ & 19.9 (28.3) & 21.6 (32.0) & 19.9 (17.0) \\
$z \sim 0.43$ & 8.4 (16.3) & 5.9 (19.5) & 13.0 (37.2) \\
$z \sim 0.35$ & 33.9 (14.6) & 54.2 (24.6) & 27.3 (33.3) \\ \hline
Subsample & ``Clean'' & Smoothed & \\
\hline
$z \sim 0.55$ & 15.8 (27.1) & 7.0 (20.8) & \\
$z \sim 0.49$ & 17.0 (30.2) & 11.0 (24.8) & \\
$z \sim 0.43$ & 6.8 (34.1) & 5.6 (30.1) & \\
$z \sim 0.35$ & 32.5 (27.3) & 12.8 (11.2) & \\ \hline
\end{tabular}
\end{center}
\end{table}

We complement the $\chi^2$ statistical analysis by applying the False
Discovery Rate (FDR \citep{Miller2001,Hopkins2002}) technique to the
combination of all of our cross--correlation functions into a single
significance test of our whole signal. The utility of the FDR
statistic over combining the individual $\chi^2$'s for different
redshift slices comes from the fact that it is conservative and
designed to work with highly correlated datasets. FDR works by
controlling $\alpha$, the fraction of false detections compared to the
total number of detections\citep{Miller2001}.  For a given $\alpha$,
the FDR technique provides a threshold above which bins are rejected
from a null hypothesis, in this case $w_{gT,i} = 0$ for all $i$
angular bins, with the guarantee that the true number of errors is
always $\le\alpha$.  Using the three highest redshift subsamples,
cross-correlated with the five CMB maps ($5\times10\times3=150$ bins
in total), we reject, with an $\alpha=0.25$, 144 bins with the
jack--knife errors and 113 bins for the random CMB map errors. Therefore, at
least 108 (84) of our 150 angular bins are rejected from the null
hypothesis using the jack--knife (random maps) errors.

\section{Theoretical Models}\label{sec:model}

We now compare our results to simple physical models of the ISW and SZ
effects. A more detailed theoretical analysis of our results will
appear in a future paper. We assume here a $\Lambda$CDM cosmology
given by the fits to the WMAPext data\citep{SpergelEtAl2003}:
$\Omega_{\rm tot}=1$, $h=0.72$, $\Omega_{\rm m}=0.29$, $\Omega_{\rm
b}=0.047$, $\sigma_8=0.9$, and $n_{\rm s}=0.99$.

We expect our observed cross-correlation functions to be the sum of
the induced correlations from the ISW and SZ
effects\citep{PeirisSpergel2000}, {\it i.e.},
\begin{equation}
w_{\rm gT}(\theta) = w_{\rm ISW}(\theta) + w_{\rm SZ}(\theta).
\end{equation}
The ISW effect is generated by the decay of gravitational potentials in an
open or dark energy dominated universe.  This decay couples to the dark matter
momentum, meaning that our galaxy-temperature cross-correlations will
be a function of the galaxy-momentum power spectrum:
\begin{eqnarray}
w_{\rm ISW}(\theta) = 
\bar{b} \int \frac{dk}{k} \int d\chi \,{\cal K}_{\rm ISW}(k,\theta,\chi)\,
D(\chi)\,P(k), \label{eq:w_isw}
\end{eqnarray}
where $\chi$ is the co-moving distance, $D(\chi)$ is the linear growth factor, 
$P(k)$ is the linear dark matter power spectra and $\bar{b}$ 
is the linear
galaxy bias at the peak of the galaxy redshift distribution ($\bar{z}$).  The 
kernel (${\cal K}_{\rm ISW}$) is given by
\begin{equation}
{\cal K}_{\rm ISW}(k,\theta,\chi) = 3 H_0^2 \Omega_{\rm m} T_0 W_{\rm g}(\chi) 
\frac{\partial}{\partial \chi} \left [ \frac{D(\chi)}{a(\chi)} \right ] 
J_0(k\theta\chi),
\end{equation}
where $H_0$ is the Hubble constant, $T_0$ is the CMB temperature today in 
$\mu$K, $W_{\rm g}$ is the normalized galaxy distribution and $J_0$ is the zero
order Bessel function.

The SZ effect is the cross-correlation of the average free-electron pressure
($\propto n_{\rm e}T_{\rm e}$) along the line of sight, with the projected
galaxy density.
The pressure-galaxy power spectrum is therefore,
\begin{equation}
w_{\rm SZ}(\theta)=\overline{n_{\rm e}T_{\rm e}b_{\rm P}b}\int dk\,k\int d\chi\,
{\cal K}_{\rm SZ}(k,\theta,\chi)\,D^2(\chi)\,P(k), \label{eq:w_sz}
\end{equation}
where $n_{\rm e}$, $T_{\rm e}$, and $b_{\rm P}$ are the electron density,
electron temperature, and pressure bias of the gas respectively, while
$\overline{\phantom{X}}$ represents the path length weighted average.  In the
Rayleigh-Jeans portion of the thermal CMB spectrum probed by WMAP, the kernel
(${\cal K}_{\rm SZ}$) is given by
\begin{equation}
{\cal K}_{\rm SZ}(k,\theta,\chi) = 
-2\frac{\sigma_{\rm T} k_{\rm B} T_0 W_{\rm g}(\chi)}{m_{\rm e}c^2}
J_0(k\theta\chi),
\end{equation}
where $\sigma_{\rm T}$ is the Thompson scattering cross-section and $k_{\rm
B}$ is Boltzmann's constant.  If we set $n_{\rm e}$ to the mean electron
density at $\bar{z}$, we have two free parameters to fit to the data, {\it
i.e.}, $\bar{b}$ and $\overline{T_{\rm e}b_{\rm P}}$.

\begin{figure}[t]
\includegraphics[width=240pt]{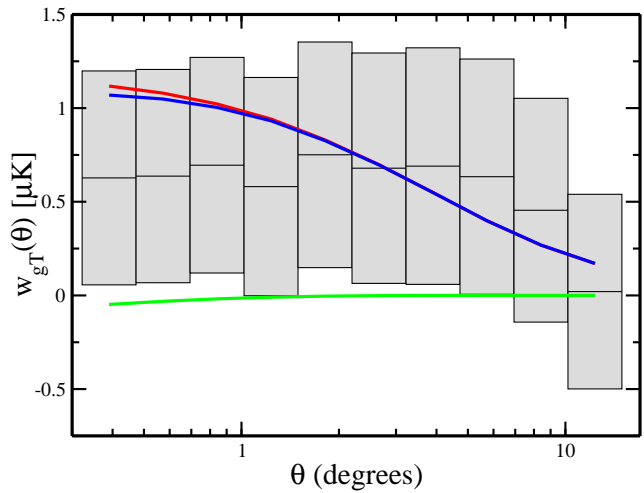}
\caption{The comparison of our theoretical predictions to our
measurement of the W-channel WMAP map cross-correlated with the $z
\sim 0.49$ galaxy subsample.  The fitted ISW cross-correlation
function is given by the red line, the SZ by the green line and the
sum of the two is given by the blue line. We find $\bar{b} = 5.47 \pm
1.82$ and $\overline{T_{\rm e} b_{\rm P}} = 0.15 \pm 0.61$, where the
errors are unmarginalized.}
\label{fig:theory_comp}
\end{figure}

Figure~\ref{fig:theory_comp} presents an example of our best fit
models for the $z \sim 0.49$ galaxy subsample cross-correlated with
the W-channel WMAP map. For this cross--correlation function, we find
that the best fit model is preferred to the null hypothesis (Section
\ref{sec:significance}) at the 99\% confidence level, based on the
observed difference between the $\chi^2$ values of the model fit
($\chi_{model}^2$) and the null hypothesis ($\chi_{null}^2$), {\it
i.e.}, $\delta \chi^2 \equiv \chi_{model}^2 - \chi_{null}^2 = 9.1$ for
2 degrees of freedom, using the jack--knife errors.  If we consider
the three highest redshift galaxy subsamples, cross--correlated with
the 5 WMAP maps (W,Q,V,''clean'',smoothed), we find that, using the
jack--knife covariance matrix ($C_J$), 14 of these 15
cross--correlation functions are better fit by our physical model than
by the null hypothesis at the $>95\%$ confidence level (based on the
$\delta \chi^2$ values). If we assume the more conservative random map
covariance matrix ($C_R$), then four of the fifteen functions are
again rejected at the same confidence (or eight at $>80\%$
confidence). In summary, a majority of our cross--correlation
functions are much better fit by our physical model than by the null
hypothesis.

We find a somewhat stronger SZ signal in the higher redshift bins,
although the SZ effect is always sub-dominant to the ISW signal.  We
also see higher values for $\bar{b}$ for the $z \sim 0.55$ galaxy
subsample than for $z \sim 0.43$ galaxy subsample, but the fit values
of $\bar{b}$ are sensitive to the details of our photometric redshift
distributions. We will present a more detailed analysis and
interpretation in a future paper.

\section{Conclusions} \label{sec:conclusions}

We present here measurements of the cross-correlation function between
the currently available SDSS galaxy data and the WMAP CMB maps of the
sky.  The individual measurements in Figure~\ref{fig:lrg_w_gt} are
significant at the $>90\%$ confidence, while FDR strongly rejects the
null hypothesis. The achromatic nature of the signal (in the Q, V, and
W frequency channels) confirms the signal is of cosmological origin.
Our results are consistent with the NVSS-CMB
correlations\citep{boughn/crittenden:2003,Nolta2003}, but may be harder to
reconcile with other optical--CMB measurements of the ISW
effect\citep{fosalba/etal:2003a,Myers2003}. These results are
consistent with the predicted ISW effect (which is largely independent
of dark energy properties) and a minor SZ contribution. Assuming a
flat universe, our detection of the ISW effect provides independent
physical evidence for the existence of dark energy.  In future papers,
analyzing all available SDSS data, we will present a more detailed
comparison between the theoretical models and the data, as well
consistency checks of the observed correlation amplitude with galaxy
biases and dark energy models.

\acknowledgments{ The authors would like to thank David Spergel for a
number of useful suggestions regarding the use of the WMAP data and
analysis.  Thanks also to Robert Crittenden, Scott Dodelson, Wayne Hu,
Lloyd Knox, Jeff Peterson, Kathy Romer, and Bhuvnesh Jain for useful
conversations and suggestions. DJE is supported by NSF AST-0098577 and
a Sloan Research Fellowship.

We acknowledge the use of 
HEALPix \citep{healpix}.

The Sloan Digital Sky Survey (SDSS) is a joint
project of The University of Chicago, Fermilab, the Institute for
Advanced Study, the Japan Participation Group, The Johns Hopkins
University, the Los Alamos National Laboratory, the
Max-Planck-Institute for Astronomy (MPIA), the Max-Planck-Institute
for Astrophysics (MPA), New Mexico State University, University of
Pittsburgh, Princeton University, the United States Naval Observatory,
and the University of Washington.

Funding for the project has been provided by the Alfred P. Sloan Foundation, 
the Participating Institutions, the National Aeronautics and Space 
Administration, the National Science Foundation, the U.S. Department of 
Energy, the Japanese Monbukagakusho, and the Max Planck Society. 
}

\def\refe {\par \hangindent=.7cm \hangafter=1 \noindent}
\def\apj { ApJ }
\def\astroph{{\tt astro-ph/}} 
\def\aap {A \& A }
\def\ajs{ ApJS }
\def\apss{ Ap\&SS }
\def\aj{AJ}
\def\prd{Phys ReV D}
\def\physrep{Phys Rep}
\def\apjs{ ApJS }
\def\mnras { MNRAS }
\def\apjl { Ap. J. Let. }

\end{document}